\documentclass{svmult}
\usepackage{mathptmx}
\usepackage{helvet}
\usepackage{courier}
\usepackage{multicol}
\usepackage{makeidx}
\usepackage{footmisc}
\usepackage{amsmath}
\usepackage{subfigure}
\usepackage{graphicx}
\makeindex
\newcommand{\boldpsi}{\boldsymbol{\psi}}

\begin{document}
\title*{An Introduction to the Density Matrix Renormalization Group Ansatz in
  Quantum Chemistry}
\author{Garnet Kin-Lic Chan, Jonathan J. Dorando, Debashree
  Ghosh, Johannes Hachmann, Eric Neuscamman, Haitao Wang, and Takeshi Yanai}
\institute{Garnet Kin-Lic Chan, Jonathan J. Dorando, Debashree Ghosh,
  Johannes Hachmann, Eric Neuscamman, and Haitao Wang  \at Department of Chemistry and Chemical Biology, Cornell
  University, Ithaca, New York 14853-1301, USA.  \and Takeshi Yanai \at
  Department of Theoretical and Computational Molecular Science,
  Institute for Molecular Science, Okazaki, Aichi 444-8585,
  Japan. \and \textit{Corresponding author}: Garnet
  Kin-Lic Chan, \email{gc238@cornell.edu}}
\authorrunning{Garnet Kin-Lic Chan et al.}
\titlerunning{An Introduction to the DMRG Ansatz in Quantum Chemistry}
\maketitle

\abstract*{The Density Matrix Renormalisation Group (DMRG) is an electronic
  structure method that has recently been applied to
  \textit{ab-initio} quantum chemistry. Even at this early stage, it has enabled the solution of
  many problems that would previously have been intractable with any other
  method, in particular, multireference problems with very large
  active spaces. Historically, the DMRG was not originally formulated from a
wavefunction perspective, but rather in a Renormalisation Group (RG)
language. However, it is now
realised that a wavefunction view of the DMRG provides a more
convenient, and in some cases more powerful, paradigm. 
Here we provide an expository introduction to the DMRG
ansatz in the context of quantum chemistry.   }

\newcommand{\etal}{{et al.}}

\section{Introduction}
The Density Matrix Renormalization Group (DMRG) is an electronic
structure method that has
recently been applied to  \textit{ab-initio} quantum chemistry. The method originated in the condensed matter community
with the pioneering  work of White
\cite{White1992,White1993}. Although the earliest quantum
chemistry implementations are only a few years old, the DMRG has already
been used to solve many problems that would have been intractable with any
other method, and especially,  multireference problems
with very large active spaces. 
For example,  we have used the DMRG
to study systems ranging from molecular potential energy
curves \cite{Chan2003,Chan2004b}, to excited states of large conjugated polymers \cite{Hachmann2007,Dorando2007}, to
metal-insulator type transitions in hydrogen chains \cite{Hachmann2006}. In each case, we
have obtained accuracies close to the (estimated) exact Complete
Active Space Configuration Interaction (CASCI) or Complete Active
Space Self-Consistent-Field (CASSCF) result, for active spaces  well outside the range of
traditional algorithms e.g. 100 active electrons in 100 active
orbitals \cite{Hachmann2006}. Unlike a traditional CAS (where the active space  wavefunction is
obtained in a brute-force Full Configuration Interaction expansion) the DMRG utilises a compact wavefunction ansatz. However, this
ansatz is very flexible,  is well-suited to nondynamic
correlation, and in the cases of long molecules, provides a near
optimal, local description of multireference correlations.

Historically, the DMRG was not originally formulated from a
wavefunction ansatz perspective, but rather in the Renormalisation Group (RG)
language of Wilson's Numerical RG \cite{Wilson1975,Wilson1983,
  White1992, White1993}, from which it is descended. The
original quantum chemical implementations of the DMRG were also
described from an RG point of view (e.g. \cite{White1999, Mitrushenkov2001,
  Chan2002, Legeza2003dyn, Moritz2007}). Although the 
mathematical form of the DMRG ansatz  has been known for some time
\cite{Fannes1992,Fannes1994, Ostlund1995,Rommer1997}, only in
recent years  has it been 
realised that the   wavefunction view of the DMRG provides a more
convenient and in many cases more powerful paradigm, 
and this has led to fundamental advances in the
DMRG method itself \cite{Verstraete2004mpdo,Verstraete2004pbc,Verstraete2004peps,Perez-Garcia2007,Schuch2007,Murg2007,Verstraete2005,White2004,Daley2004,Vidal2004,Vidal2006,Hallberg2003,Hallberg2006,Schollwock2005,Hachmann2006}.

The current article provides an expository introduction to the DMRG
in quantum chemistry from the wavefunction point of view. This  is
 complementary to earlier articles that use the RG based formulation and the
first-time reader will benefit from reading such articles
alongside the current one. It is not our intention to provide a comprehensive
review of the DMRG method even within the restricted domain of
quantum chemistry. Thus we do not  pretend to 
 survey the literature except to  say at the start that the field of quantum chemical
DMRG has  developed through the work of White 
\etal  \cite{White1999,Daul2000,Rissler2006},  Mitrushenkov \etal  \cite{Mitrushenkov2001,Mitrushenkov2003,Mitrushenkov2003nort}, our
contributions \cite{Chan2002,Chan2003,Chan2004,Chan2004b,Chan2005,Hachmann2006,Dorando2007,Hachmann2007}, the work of Legeza,
Hess \etal  \cite{Legeza2003dyn,Legeza2003qie,Legeza2003lif,Legeza2004}, the work of Reiher \etal  \cite{Moritz2005orb,Moritz2005rel,Moritz2006,Moritz2007}, and most recently the
work of Zgid
and Nooijen \cite{Zgid2008}. Also related, but too numerous to cite
in full here, are the developments   with semi-empirical
Hamiltonians; some representative early works are those in
\cite{Ramasesha1997,Yaron1998,Shuai1998,Fano1998,Bendazzoli1999,Raghu2002a,Raghu2002b}.
In addition, we mention
again that the DMRG has its origins
in the  condensed matter community and thus 
excellent sources of information which provide this
perspective  are the recent reviews of Schollw\"ock \cite{Schollwock2005} and
Hallberg \cite{Hallberg2003,Hallberg2006}. 

The structure of our article is as follows. We begin by introducing
the underlying DMRG ansatz and examining some of its special
properties in sections \ref{sec:ansatz1} and \ref{sec:properties}.
In sections \ref{sec:rg1} and
\ref{sec:rg2} we explain the connection between the wavefunction ansatz,
and the original Renormalisation Group language within which the
DMRG is usually described. In section \ref{sec:matrix} we describe how
the structure of the DMRG wavefunction allows the efficient evaluation
of Hamiltonian matrix elements.  Finally, we finish with some brief thoughts and conclusions in
section \ref{sec:conclusions}.

\section{Motivation for the DMRG Ansatz}

\label{sec:ansatz1}
The primary challenge in quantum chemistry is to find a good
approximation to the electronic wavefunction of a quantum state. We can
express any $N$-electron wavefunction in a complete basis of Slater
determinants, through the  Full Configuration Interaction (FCI) expansion,
\begin{align}
|\Psi\rangle &= \sum_{n_1n_2n_3 \ldots n_k} \Psi^{n_1n_2n_3 \ldots n_k}
|n_1 n_2 n_3 \ldots n_k\rangle, \\
\{ n_i \} & = \{ |0\rangle, |1^\alpha\rangle, |1^\beta\rangle,
|2^{\alpha\beta}\rangle \}, \\
 \sum_{i} n_i & = N.
\end{align}
Here $|n_1 \ldots n_k\rangle$ is the occupation number representation
of the Slater determinant
where  $n_i$ is the occupation of site (i.e. orbital) $i$. The total number of
orbitals is $k$  and $N$ is the total number of electrons. 

The dimension of the coefficient tensor $\Psi$
in the above expansion is $4^k$, which is
intractable  for  values of $k$ much larger than 10. Therefore, we would like to 
 find an ansatz where $\Psi$ is expressed more compactly. In particular, we would want such an ansatz to require only a
\textit{polynomial} amount of information as a function of the number
of orbitals in the system, $k$.

A very simple  ansatz would be to approximate the high-dimensional
coefficient tensor $\Psi$ by
a tensor product of vectors $\psi^1 \ldots \psi^k$, which we shall
call site functions, 
\begin{align}
\Psi \approx \psi^1 \otimes \psi^2 \otimes \psi^3 \ldots \otimes
\psi^k.
\end{align}
Using the notation $\psi^{n_1}$ to denote the $n$th element of
$\psi^1$, i.e. $\psi^{n_1} = \psi^i_n$, we can also write
\begin{align}
\Psi^{n_1n_2n_3\ldots n_k} \approx \psi^{n_1} \psi^{n_2} \psi^{n_3} \ldots \psi^{n_k}.
\end{align}
Note that each site function $\psi$ is \textit{not} an orbital but rather a vector of length 4, and  $\psi^{n_1},
\psi^{n_2}$ represent elements of the different vectors $\psi^1, \psi^2$.
This ansatz contains only $4k$ parameters and is certainly tractable. However,
it is also not, in general, very accurate. So, let us try to improve the ansatz by
increasing the flexibility of the site functions $\psi$. 
We can  introduce additional \textit{auxiliary}
indices, i.e. 
\begin{equation}
\psi^{n_p} \to \psi^{n_p}_{ii^\prime}.
\end{equation}
The new  indices $i,i^\prime$
are auxiliary in the sense that they do not appear in the
final coefficient tensor $\Psi$ and must be contracted over in
some fashion. The simplest arrangement is to contract the indices
sequentially from one $\psi$ site function to the next, i.e.
\begin{equation}
\Psi^{n_1n_2n_3\ldots n_k} \approx \sum_{i_1i_2i_3 \ldots i_{k-1}} \psi^{n_1}_{i_1}
\psi^{n_2}_{i_1 i_2} \psi^{n_3}_{i_2 i_3} \ldots \psi^{n_k}_{i_{k-1}}. \label{eq:mps}
\end{equation}
For simplicity, we will assume that the dimensions of all
auxiliary indices are chosen to be the same, and we shall call this
dimension $M$. Then each site function $\psi$ is  a
3-tensor of dimension $4 \times M \times M$, and the total
number of parameters in the  wavefunction ansatz is $4M^2 k$. 

This is, in essence, the DMRG ansatz for $M$
states. (More precisely, it is the ansatz used in the
one-site DMRG algorithm, as explained later). Note that
by increasing the dimension $M$, we can make the  approximation arbitrarily
exact. Because (for given $n_1 \ldots n_k$) the contraction
in Eq. (\ref{eq:mps}) is a series of matrix products, this ansatz is
referred to in the literature as the Matrix
Product State \cite{Fannes1992,Fannes1994,Ostlund1995,Rommer1997,Dukelsky1998,Verstraete2004mpdo,Verstraete2005,Verstraete2006,Perez-Garcia2007mps}. Combining the site functions explicitly with the  Slater
determinants we have
\begin{equation}
|\Psi_\text{DMRG}\rangle = \mathop{\sum_{n_1 n_2 n_3
    \ldots n_k}}_{i_1i_2i_3 \ldots i_{k-1}} \psi^{n_1}_{i_1}
\psi^{n_2}_{i_1 i_2} \psi^{n_3}_{i_2 i_3} \ldots \psi^{n_k}_{i_{k-1}} |n_1
n_2 n_3 \ldots n_k\rangle. \label{eq:dmrg_ansatz}
\end{equation} 

Before continuing, let us first establish some notation. The
site functions $\psi$ in Eq. (\ref{eq:dmrg_ansatz}) are 3-tensors. However, the  notation of
linear algebra is designed primarily for  vectors (1-tensors) and
matrices (2-tensors). 
Naturally, any 3-tensor
can  be considered as an array of matrices, so long as we specify which
two indices are the matrix indices and which is the 3rd (array)
index. When viewing the site function as an
array of matrices, we  will write the 3rd (array) index on the top. Thus in
this notation, we have
\begin{align}
\text{Matrix} & : [\boldpsi^{n_p}] \ \text{(dimension $M\times M$)}\nonumber\\
\text{Elements} & : \psi^{n_p}_{i_{p-1} i_p}
\end{align}
and the DMRG wavefunction (\ref{eq:dmrg_ansatz}) is written as
\begin{equation}
|\Psi_\text{DMRG}\rangle = \mathop{\sum_{n_1 n_2 n_3
    \ldots n_k}} [\boldpsi^{n_1}]
[\boldpsi^{n_2}] [\boldpsi^{n_3}] \ldots [\boldpsi^{n_k}] |n_1
n_2 n_3 \ldots n_k\rangle
\end{equation} 
(Note that the first and last site functions $[\psi^{n_1}],
[\psi^{n_k}]$ have dimensions $1\times M$ and $M \times 1$ respectively).

Alternatively, we can view a 3-tensor as a single matrix if we group
two  indices together to make a compound index. This view will be
useful when discussing the
renormalised basis and canonical representations of the DMRG
wavefunction in sections \ref{sec:rg1} and \ref{sec:rg2}. Depending on
the context, we will either group the $n$ index with the left or the
right auxiliary indices, giving
\begin{align}
\text{Matrix} & : [\boldpsi^{p}] \ \text{(dimension $4M \times M$)} \nonumber\\
\text{Elements} & : \psi^p_{ni, i^\prime} \nonumber \\
\text{\textit{or}} \ \ \ \text{Matrix} & : [\boldpsi^{p}] \ \text{(dimension $M \times 4M$)} \nonumber\\
\text{Elements} & : \psi^p_{i, ni^\prime}  \label{eq:renorm_form}
\end{align}
Note that  the superscript $p$  here denotes the $p$th site function in
the DMRG ansatz (\ref{eq:dmrg_ansatz}),  not any particular element of the  site function.

\section{Properties of the DMRG ansatz}

\label{sec:properties}
Let us now examine some properties of  the DMRG ansatz.
\begin{enumerate}
\item {\textit{Variational}}: Since we have an explicit wavefunction, the
  expectation value of the energy provides a variational upper bound
  to the true energy and in practice  DMRG energies are evaluated in
  this way. As $M$ is increased, the DMRG energy converges from above to the exact
  energy.
\item {\textit{Multireference}}: There is no division into occupied and virtual
  orbitals, all orbitals appear on an equal
  footing in the  ansatz (\ref{eq:dmrg_ansatz}). In particular,  the Hartree-Fock reference has no
special significance here. For this reason, we expect (and observe) the ansatz
   to be  very well-balanced for describing nondynamic correlation in
  multireference   problems (see e.g. \cite{Chan2004, Chan2004b, Hachmann2006}). Conversely, the ansatz is 
  inefficient for describing \textit{dynamic} correlation, since this
  benefits from knowledge of the  occupied and virtual spaces.
\item {\textit{Size-consistency}}: The
  DMRG ansatz is size-consistent within  a localised basis. Consider 
a system $AB$ composed of two spatially separated, non-interacting
  subsystems $A$ and $B$. 
Associate localised orbitals $1 \ldots a$
  with subsystem $A$ and $a+1 \ldots a+b$ with subsystem
  $B$. Then, the DMRG wavefunction for $AB$  factorises into a product of
  DMRG wavefunctions for $A$ and $B$. First expand the DMRG wavefunction
\begin{align} 
 |\Psi^\text{AB}_\text{DMRG}\rangle& = \mathop{\sum_{n_1 \ldots n_{a+b}}}_{i_1 \ldots i_{a+b-1}}
 \psi^{n_1}_{i_1}
 \ldots \psi^{n_{a}}_{i_{a-1} i_a} \psi^{n_{a+1}}_{i_a i_{a+1}}
   \ldots \psi^{n_{a+b}}_{i_{a+b-1}} |n_1 \ldots
   n_{a} n_{a+1} n_{a+b}\rangle \nonumber\\
 &=\sum_{i_{a}} 
\Bigl(
\mathop{\sum_{n_1 \ldots n_{a}}}_{i_1 \ldots
 i_{a-1}} 
 \psi^{n_1}_{i_1} \ldots \psi^{n_{a}}_{i_{a-1} {i_{a}}} |n_1 \ldots
 n_{a}\rangle
\nonumber \\
& \times 
 \mathop{\sum_{n_{a+1} \ldots n_{a+b}}}_{i_{a+1} \ldots
 i_{a+b-1}}
  \psi^{n_{a+1}}_{i_{a} i_{a+1}} \psi^{n_{a+b}}_{i_{a+b-1}}
    |n_{a+1} \ldots n_{a+b}\rangle
{\Bigr) }. \label{eq:partial_sep}
\end{align}
Then note that we can write a separable wavefunction $|\Psi^{\text{AB}}\rangle
= |\Psi^\text{A}\rangle |\Psi^\text{B}\rangle$ formally as $|\Psi^\text{AB}\rangle = \sum_{i=1}^1
|\Psi^\text{A}_i\rangle |\Psi^\text{B}_i\rangle$ and thus we can take
the dimension of index $i_a$  which
couples systems $A$ and $B$ above to be 1, giving
\begin{align}
|\Psi^\text{AB}_\text{DMRG}\rangle &=\mathop{\sum_{n_1 \ldots n_{a}}}_{i_1 \ldots
 i_{a-1}} 
 \psi^{n_1}_{i_1} \ldots \psi^{n_{a}}_{i_{a-1}} |n_1 \ldots
 n_{a}\rangle
 \mathop{\sum_{n_{a+1} \ldots n_{a+b}}}_{i_{a+1} \ldots
 i_{a+b}}
  \psi^{n_{a+1}}_{i_{a+1}} \psi^{n_{a+b}}_{i_{a+b-1}}
    |n_{a+1} \ldots n_{a+b}\rangle \nonumber\\
&=|\Psi^\text{A}_\text{DMRG}\rangle |\Psi^\text{B}_\text{DMRG}\rangle .
\end{align}

\item{\textit{Compactness and efficiency of the ansatz}}: The number of variational parameters
  in  the DMRG ansatz is $O(M^2 k)$. How large do we need $M$ to be to achieve
  a good accuracy? If we choose, for a given index $i_p$, $M=1$, then the wavefunction
  factorises into a simple product of contributions from the spaces
  $\{ n_1 \ldots n_p \}$ and $\{n_{p+1} \ldots n_k\}$. Increasing $M$ then
  introduces additional correlations or entanglement between the
  wavefunction components in the two spaces. The  $M$ required for a
  given accuracy thus depends on the correlations in the specific
  state of the molecule. However,  we have 
seen in  our applications that for appropriate problems, even modest
  $M=O(100-1000)$  can allow us to obtain 
  very good accuracy and to solve problems that are insoluble  with other techniques.
Of course, having a small number of variational parameters does not
  guarantee that an ansatz can be manipulated
 efficiently. (Witness the difficulty in evaluating the variational
 energy corresponding to a Coupled Cluster wavefunction!)
As we shall see in Sec. \ref{sec:matrix}, the product
  structure of the DMRG ansatz enables matrix elements to be
  evaluated without ever reconstructing the DMRG  coefficients
 in the full Slater determinant expansion, thus bypassing the exponential complexity. (Although one can do so  if one
  wishes, e.g. for the purposes of analysing the DMRG
 wavefunction, as  in \cite{Moritz2007}). Finally, we note that the DMRG  incorporates correlations between orbital spaces
 in a sequential  manner, i.e. the first set of auxiliary
indices $i_1$ entangles  spaces $\{n_1\}$ and
$\{n_2 \ldots n_k\}$,  $i_2$  entangles
spaces $\{n_1n_2\}$ and $\{n_3 \ldots n_k\}$ and so on. For this
 reason, the DMRG ansatz  performs best if strongly-correlated orbitals are placed
  next to each other in the ansatz \cite{Chan2002, Legeza2003qie,
 Moritz2005orb, Rissler2006}.

\item{\textit{A local multireference ansatz for long molecules}}:
The DMRG wavefunction is
particularly well-suited to long molecules where it can be viewed as a naturally local
multireference ansatz. In long molecules (i.e. those where
one of the dimensions is much larger than
the other two)  with a finite electronic correlation length,
 we can divide the molecule at any point
 along the backbone and expect the degree of entanglement between the two
 resulting subsystems to be independent of the point of division and
 the length of the chain.
 Thus, for such problems,  the $M$ required for a  given accuracy is \textit{independent} of the length of the
system and the number of variational parameters in the DMRG
wavefunction is simply $\text{const} \times O(k)$, as should be in a
local ansatz. However, unlike in other
local correlation approaches the DMRG provides a local
\textit{multireference} ansatz. It is  this local nature  even in the
presence of strong nondynamic correlations which has allowed us to
solve very large active space multireference correlation problems in long molecules
\cite{Hachmann2006,Dorando2007,Hachmann2007}.


\begin{figure}[t]
\centering
\subfigure[DMRG] 
{
    \includegraphics[width=7cm]{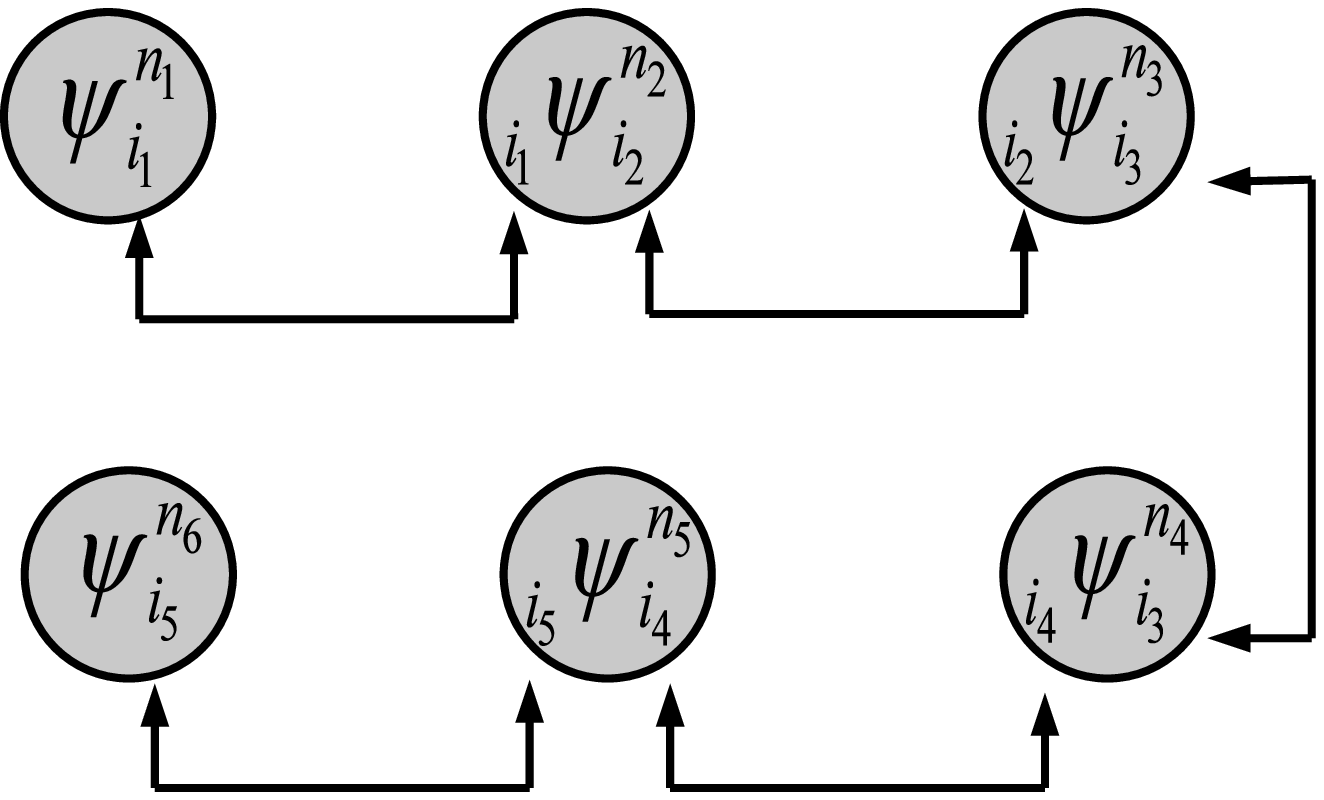}
}
\\
\subfigure[PEPS]
{
    \includegraphics[width=7cm]{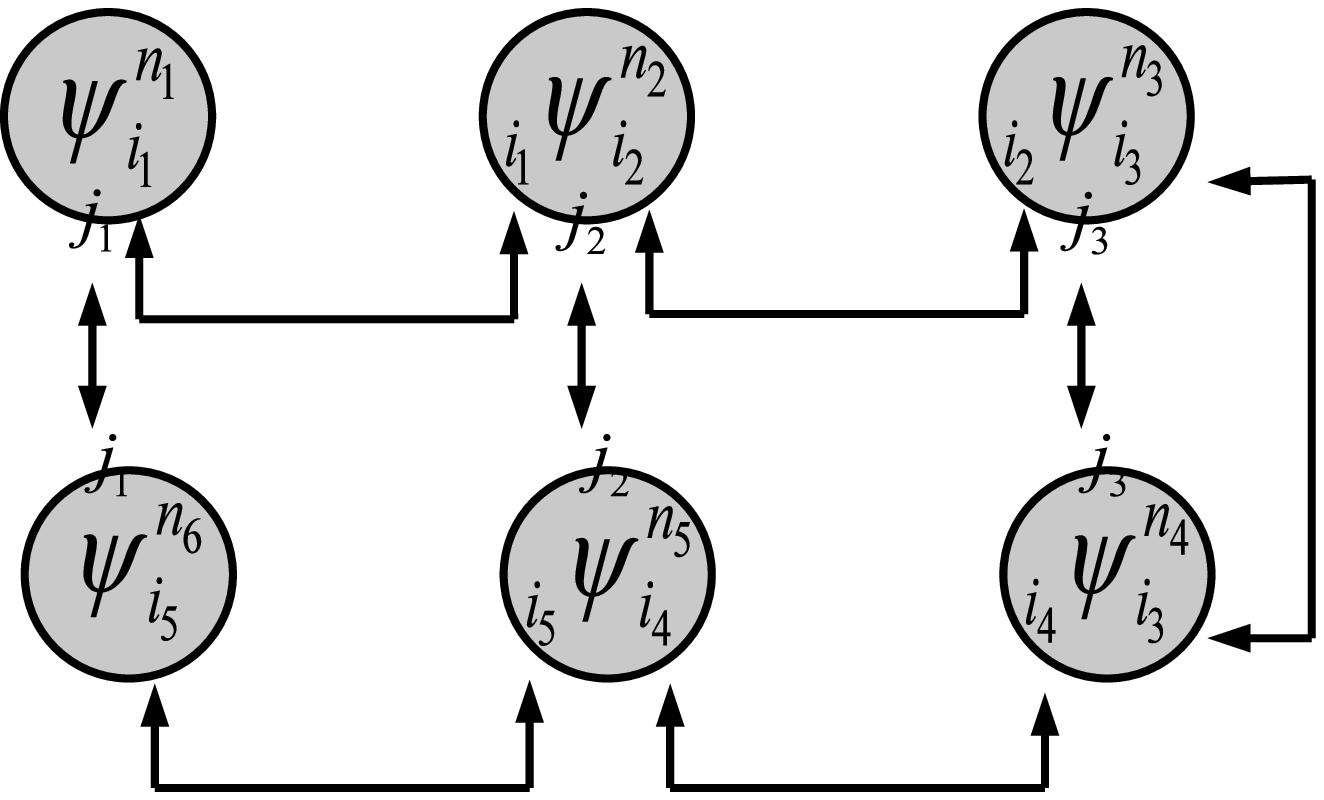}
}
\caption{\label{fig:peps} Density Matrix Renormalisation Group  and Pair Entangled Product State wavefunctions for
two rows of atoms. Note in the DMRG ansatz, the site functions are
coupled sequentially, which prevents the efficient description of
correlations between the rows. However, in the PEPS ansatz, addition
indices are added to the site functions (e.g. $\psi^{n_2}_{i_1 i_2}
\to \psi^{n_2}_{i_1 i_2 j_2}$) whose coupling  directly captures the
inter-row correlations.}
\end{figure}

In problems which are large in two or three dimensions, the degree of
entanglement between two subsystems grows exponentially with the length
of the border, and thus the preceding considerations no longer apply. 
We might then ask,  can we modify the DMRG ansatz
to obtain a naturally local multireference description  for large systems with arbitrary dimensionality? Recently, this has been shown to be possible. Consider, for
example, two rows of atoms (each with one localised orbital) arranged
as in Fig. \ref{fig:peps}. The first sub-figure illustrates the sequential coupling between
 orbital spaces that is contained in  the DMRG wavefunction, which  is inefficient at describing correlations
 between atoms in different rows. In the second sub-figure, however, 
we have added additional auxiliary indices to couple
the site functions both along the rows as well as along the columns in a non-sequential manner. This is
the basis for the so-called Pair-Entangled Product State wavefunctions
which present one of the most promising new developments in this
area \cite{Verstraete2004peps,Perez-Garcia2007,Schuch2007}. 
\end{enumerate}

\section{The Renormalized Basis}

\label{sec:rg1}
As we have discussed above, the auxiliary indices of the site functions introduce couplings between
the orbital spaces in the DMRG ansatz. In addition, they can also be
provided with a direct physical interpretation.
Just as the index $n_i$ is associated with the Fock space  of orbital $i$, so can we also
associate a set of \textit{renormalised many-body
spaces} with the auxiliary indices of each site function $\psi$. This
provides the Renormalisation Group (RG) interpretation of the DMRG
wavefunction. Consider, for example, the first set of auxiliary indices $i_1$. We first
 perform the summation in the DMRG wavefunction
 expression over $n_1$, which couples $\psi^{n_1}_{i_1}$ with the set of
 states $\{ |n_1\rangle \} = \{ |0\rangle, |1^\alpha\rangle,
 |1^\beta\rangle, |2^{\alpha\beta}\rangle\}$. This formally defines a space
 $\{ i_1 \}$ with basis functions $|i_1\rangle$ 
\begin{equation}
|i_1\rangle = \sum_{n_1} \psi^{n_1}_{i_1} |n_1\rangle
\end{equation} 
or more succinctly
\begin{equation}
 \{ i_1 \} = \hat{\psi}^1 \cdot \{ n_1 \}. 
\end{equation}
Of course, the  transformation of the $\{n_1\}$ orbital Fock space  by the $\psi^1$
site function  is  trivial.  (Indeed, if, as is usual, we do not
allow $\psi$ to mix states with
different particle numbers or spin, we would simply have $|i_1\rangle =
|n_1\rangle$ for all 4 states).
However, things are more interesting, when we consider the spaces associated with
later sets of auxiliary indices. For example, repeating the above
exercise for $i_2$
\begin{align}
|i_2\rangle &= \mathop{\sum_{n_1n_2}}_{i_1} \psi^{n_1}_{i_1} \psi^{n_2}_{i_1i_2} |n_1
n_2\rangle  \\
&= \mathop{\sum_{n_2}}_{i_1} \psi^{n_2}_{i_1i_2} |i_1n_2\rangle , \\
\{ i_2 \} &= \hat{\psi}^2  \cdot  \{ i_1 n_2 \} = \hat{\psi}^2  \cdot \hat{\psi}^1 \cdot \{ n_1 n_2 \} .
\end{align}
In general for the space $\{ i_p \}$ and the associated
basis $|i_p\rangle$, we write
\begin{align}
\{ i_p \} &= \hat{\psi}^p \cdot \{ i_{p-1} n_p \} \nonumber \\
          &= \hat{\psi}^p \cdot \hat{\psi}^{p-1} \cdots \hat{\psi}^1 \cdot \{ n_1 n_2 \ldots
          n_p \}, \\
|i_p\rangle& = \mathop{\sum_{n_p}}_{i_{p-1}} \psi^{n_p}_{i_{p-1}i_{p}} |i_{p-1}n_p\rangle
          \nonumber\\
 &= \mathop{\sum_{n_1 \ldots n_p}}_{i_1 \ldots i_{p-1}}
          \psi^{n_1}_{i_1} \psi^{n_2}_{i_1 i_2}  \ldots
          \psi^{n_{p-1}}_{i_{p-2} i_{p-1}}\psi^{n_p}_{i_{p-1}i_p} 
          |n_1 n_2 \ldots n_p\rangle\label{eq:basis_ip}. 
\end{align}
Note that the matrix representation of $\hat{\psi}^p$ is simply the
matrix form of the site function $[\psi^p]$ described in
Eq. (\ref{eq:renorm_form}),   i.e.
\begin{align}
\langle i_p | \hat{\psi}^p | i_{p-1} n_p\rangle = \psi^p_{i_p, i_{p-1} n_1}
\end{align}
and thus we can also write Eq. (\ref{eq:basis_ip}) as
\begin{align}
|i_p\rangle &= \sum_{n_1 \ldots n_p}
          [\psi^{n_1}] [\psi^{n_2}]  \ldots
          [\psi^{n_{p-1}}] [\psi^{n_p}]
          |n_1 n_2 \ldots n_p\rangle . 
\end{align}

Now  the dimension of the  $i_p$ index  and $\{i_p\}$ space is fixed
to be at most  $M$ in the
original ansatz (\ref{eq:dmrg_ansatz}). Thus, the
action of $\hat{\psi}^p \cdots \hat{\psi}^1$ is a projective transformation from the full many-body
space down into a \textit{renormalised} many-body space of $M$
basis states, where  each  basis state $|i_p\rangle$ is expressed as  a linear
combination of many product functions $|n_1 \ldots n_p\rangle$ with
coefficients given by Eq. (\ref{eq:basis_ip}). The renormalised spaces
have a  recursive structure: $\{
i_p \}$ is obtained from $\{ i_{p-1}\}$ which is obtained from
$\{ i_{p-2}\}$ and so on.

\begin{figure}[t]
\centering
\subfigure[One site ansatz] 
{
    \includegraphics[width=7cm]{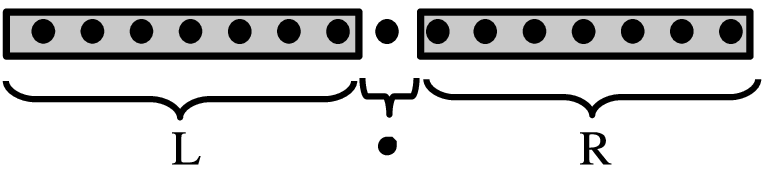}
}
\\
\subfigure[Two site ansatz]
{
    \includegraphics[width=7cm]{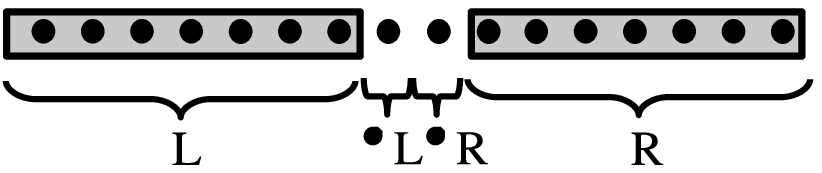}
}
\caption{\label{fig:blocking} Block diagrams for the one and two site DMRG ansaetze. }

\end{figure}

The construction of one  renormalised space from the
previous one may be
considered to proceed in two stages. To
construct the space $\{ i_p \}$, first we form the product space
\begin{equation}
\{ i_{p-1} \} \otimes \{ n_p \} \to \{ i_{p-1} n_p \}
\end{equation}
and then we apply the projective transformation
\begin{equation}
\hat{\psi}^p \cdot \{ i_{p-1} n_p \} \to \{ i_p \}.
\end{equation}
The first step is called ``blocking'' and the second step
``decimation'' in the traditional language of the Renormalisation
Group, and therein lies the basic connection between 
the DMRG  ansatz and its   RG interpretation.
It is common to represent these blocking and decimation steps in the
pictorial fashion  shown in Fig. \ref{fig:blocking}. 

\section{The Canonical Representation and Sweep algorithm}

\label{sec:rg2}
\newcommand{\bpsi}{\boldsymbol{\psi}}
\newcommand{\bham}{\boldsymbol{H}}
\newcommand{\boldV}{\boldsymbol{V}}
\newcommand{\bone}{\boldsymbol{1}}

\newcommand{\boldC}{\boldsymbol{C}}
\newcommand{\boldl}{\boldsymbol{l}}
\newcommand{\boldGamma}{\boldsymbol{\Gamma}}

The DMRG wavefunction is invariant to a class of
transformations of  the site functions $\psi$, since the associated
nested many-body spaces $\{ i_p \}$ are themselves invariant with respect to
transformations  within each space. 
The original  DMRG algorithm, which was formulated in the language of
orthogonal projective transformations following Wilson's Numerical
Renormalisation Group, in fact corresponds to particular choices of
representation of the site functions within the above invariant
class. We shall call such representations ``canonical
representations''. All existing DMRG implementations in
quantum chemistry work with  canonical representations of the DMRG
wavefunction. In addition, the use of canonical representations is closely linked with the  density
matrix interpretation of the DMRG and also with the DMRG sweep algorithm, which
provides a natural algorithm to optimise the DMRG wavefunction.


Associated with each DMRG wavefunction $\Psi$
there are $k$ canonical representations, one for each site. 
At site $p$, the canonical representation is written as
\begin{align}
|\Psi \rangle& = \sum_{n_1 \ldots n_p \ldots n_k} [L^{n_1}] \ldots [L^{n_{p-1}}]
 [C^{n_p}] [R^{n_{p+1}}] \ldots [R^{n_k}] |n_1 \ldots n_p
 \ldots n_k\rangle \\
&=\mathop{\sum_{n_1 \ldots n_p \ldots n_k}}_{l_1
 \ldots l_{p-1}, r_p \ldots r_{k-1}} L^{n_1}_{l_1} \ldots L^{n_{p-1}}_{l_{p-2} l_{p-1}}
 C^{n_p}_{l_{p-1} r_p} R^{n_{p+1}}_{r_p r_{p+1}} \ldots R^{n_k}_{r_{k-1}} |n_1 \ldots n_p
 \ldots n_k\rangle . \label{eq:dmrg_mps}
\end{align} 
Here, the site functions to the left of $p$ have been given the symbol
$L$, while those to the right have been given the symbol $R$. The $L$
and $R$ site functions, which are in this context usually called
transformation
matrices, are each orthogonal matrices when written in the
matrix representation of Eq. (\ref{eq:renorm_form}). We interpret the $L$
site functions as matrices by grouping the $n$ index with the first
auxiliary index, 
\begin{align}
(q<p: L^q_{ln,l^\prime} := L^{n_q}_{l l^\prime})
\end{align}
and in this form we have
 \begin{align}
[L^q]^T [L^q] &= [1], \\
\sum_{ln}
   L^q_{ln,l^\prime} L^q_{ln, l^{\prime\prime}}& = \delta_{l^\prime
     l^{\prime\prime}} . 
     \label{eq:l_ortho}
 \end{align}
For the $R$ site functions, we group the $n$ index with the
second auxiliary index
\begin{align}
(q>p: R^q_{r^\prime,rn}:=R^{n_q}_{r^\prime r})
\end{align}
and in this form we have
\begin{align}
[R^q] [R^q]^T& = [1], \\
\sum_{rn} R^q_{r^\prime,rn}
     R^q_{r^{\prime\prime},rn}& = \delta_{r^\prime r^{\prime\prime}}.
     \label{eq:r_ortho}
\end{align}

The $L$ and $R$ matrices each define a set of orthogonal projective
transformations, which give rise, respectively, to   two sets of 
renormalised spaces $\{ l\}$ and $\{ r\}$ associated with the
site $p$ representation of the DMRG wavefunction. 
The $\{l \}$ spaces,  $\{l_1\}, \{l_2\} \ldots$ are
built up  by incorporating the orbitals in the order $1,2
\ldots p$, 
\begin{align}
  (q<p): \{ l_q \} & = \hat{L}^q \cdot \{ l_{q-1} n_q \} \nonumber\\
                    & = \hat{L}^q \cdot \hat{L}^{q-1} \cdot \{ l_{q-2} n_{q-1} n_q
                    \}\nonumber\\
                    & = \hat{L}^q \cdot \hat{L}^{q-1} \cdots \hat{L}^1 \cdot \{ n_1
                    \ldots n_q \} 
\end{align}
and the $|l\rangle$ functions form an orthogonal renormalised basis (from the
orthogonal nature of the $[L]$ transformation matrices) for each
$\{ l \}$ space
\begin{align}
|l_q\rangle   &= \sum_{n_1 \ldots n_p} [L^{n_1}] [L^{n_2}] \ldots
                    [L^{n_{q-1}}]  [L^{n_q}]
          |n_1 n_2 \ldots n_1\rangle \label{eq:left_transform_ip},\\
          \langle l_q | l^\prime_q\rangle& = \delta_{ll^\prime}.
\end{align}

The $\{r\}$ spaces and $|r\rangle$ basis functions are defined
similarly, but now the orbitals are incorporated ``backwards'' in the order $k, k-1 \ldots p+1$
\begin{align}
(q>p): \{ r_q \}  &= \hat{R}^q \cdot \{ n_{q} r_{q+1}  \} \nonumber\\
                  &= \hat{R}^q\cdot \hat{R}^{q+1} \cdot \{ n_q n_{q+1} r_{q+2} \}
                    \nonumber\\
                  &= \hat{R}^q\cdot \hat{R}^{q+1} \cdots \hat{R}^k \cdot \{ n_q
                    \ldots n_k \},    \\
|r_q\rangle   &= \sum_{n_q \ldots n_k} [R^{n_q}] [R^{n_{q+1}}] \ldots
                    [R^{n_{k-1}}]  [R^{n_k}]
          |n_q n_{q+1} \ldots n_k\rangle\label{eq:right_transform_ip},
                    \\
     \langle r_q|r^\prime_q\rangle& = \delta_{rr^\prime}.
\end{align}

Having defined the renormalised spaces, we now see that the $C^p$ site function gives the wavefunction coefficients in the
 product space formed from  the
renormalised left basis $\{ l_{p-1}\}$, the orbital space $\{ n_p\}$,
and the renormalised right basis $\{ r_p\}$
\begin{align}
|\Psi \rangle = \sum_{lnr} C^p_{lnr} |l_{p-1} n_p r_p\rangle \label{eq:blockwf}
\end{align} 
where we have used the notation  $C^p_{lnr}:= C^{n_p}_{l_{p-1} r_p}$.

We now consider the DMRG wavefunction expressed in the canonical
representations of  sites other than $p$. Since the same wavefunction is
simply being expressed in a different representation,  this implies a
relationship between the wavefunction coefficients $C$ and
transformation matrices $L,
R$ at different sites.
Comparing representations at sites $p$, $p+1$ we see
\begin{align}
|\Psi \rangle& = \sum_{n_1 \ldots n_p \ldots n_k}
[L^{n_1}] \ldots [L^{n_{p-1}}] 
[C^{n_p}] 
[R^{n_{p+1}}]
[R^{n_{p+2}}]\ldots [R^{n_k}]
|n_1 \ldots n_p
 \ldots n_k\rangle\\
& = \sum_{n_1 \ldots n_p \ldots n_k}
[L^{n_1}] \ldots [L^{n_{p-1}}]
[L^{n_{p}}]
 [C^{n_{p+1}}]
[R^{n_{p+2}}] \ldots [R^{n_k}]
|n_1 \ldots n_p  \ldots n_k\rangle .
\end{align}
This implies
\begin{align}
[C^{n_p}] [R^{n_{p+1}}]=[L^{n_{p}}] [C^{n_{p+1}}]
\end{align}
or, switching to the alternative matrix interpretation of Eq. (\ref{eq:renorm_form})  for
$C^p, C^{p+1}$ and likewise for
$L^p, R^{p+1}$
\begin{align}
\sum_{r} C^p_{ln,r} R^{p+1}_{r, r^\prime n} = \sum_{l^\prime} L^{p}_{ln, l^\prime}
C^{p+1}_{l^\prime, n^\prime r^\prime} .
\end{align}
From $C^p$, we can determine the quantities in the
site $p+1$ canonical form that do not explicitly appear in the site
$p$ canonical form, namely $C^{p+1}, L^p$, by the singular value
decomposition (SVD) of
$C^p$,
\begin{align}
C^p_{ln,r} &= \sum_{l^\prime} L^p_{ln, l^\prime} \sigma_{l^\prime}
V_{l^\prime r} \label{eq:svd_relation},\\
C^{p+1}_{l,nr} &= \sum_{ r^\prime} \sigma_{l} V_{l r^\prime}
R^{p+1}_{r^\prime, rn}.
\end{align}

\newcommand{\lpp}{l^\prime}
\newcommand{\npp}{n^\prime}
\newcommand{\rpp}{r^\prime}

\begin{figure}[t]
\centering
\includegraphics[width=7cm]{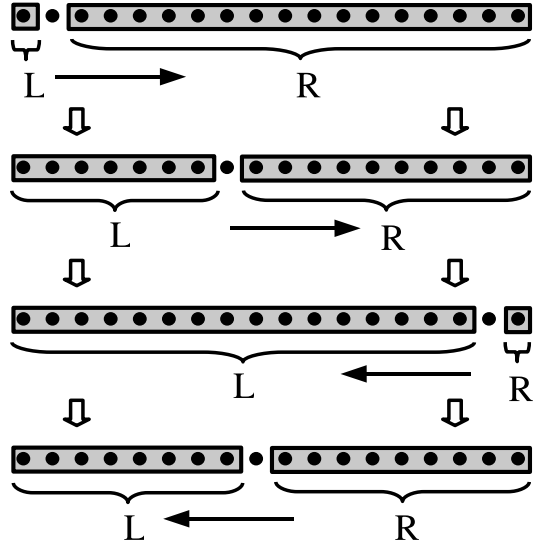}
\caption{\label{fig:sweep} The DMRG sweep algorithm with the one site
  ansatz. After all $L$ blocks are constructed going from $L \to R$,
  the sweep direction is reversed. }
\end{figure}

The connection through the SVD between the representations at
different sites leads to the density matrix formulation of the
DMRG. Recall that the
singular vectors of a matrix $M$
 may be related to the eigenvectors of 
 $M^T M$ and $MM^T$. Thus from $C^p$, we can define a symmetric object
$[\Gamma^p]= [C^p]^T [C^p]$, i.e.
\begin{align}
\Gamma^p_{ln,l^\prime n^\prime} = \sum_r C^p_{lnr} C^p_{l^\prime
  n^\prime r}.
\end{align}
$\Gamma^p$ is none other than the \textit{density matrix} associated
with the left subsystem, or ``block'' of
orbitals $1 \ldots p$, and  the  left transformation matrix $L^p$  is
obtained as the matrix of $M$ eigenvectors
\begin{align}
 \sum_{l^\prime n^\prime} \Gamma^p_{ln, l^\prime n^\prime} L^p_{l^\prime
   n^\prime, l^{\prime\prime}} = L^p_{ln, l^{\prime\prime}}
   \sigma^2_{l^{\prime\prime}} . \label{eq:dm_eigenvectors}
\end{align}
This corresponds to the traditional density matrix interpretation of the DMRG: to
obtain the canonical representation at a new site requires a basis
change into the eigenvectors of the subsystem density matrix.

The sequential set of
transformations from representation to representation along the sites
also  yields  a natural optimisation procedure
for the  DMRG wavefunction known as the \textit{sweep
algorithm}. At each site $p$, we solve the Schr\"odinger equation in
the basis $ \{ l_{p-1} n_p r_p \}$ to obtain the
coefficient matrix $C^p$, thus (dropping the subscripts on the basis
functions for simplicity)
\begin{align}
\langle l^\prime n^\prime r^\prime|\hat{H}-E |\Psi\rangle &= 0, \\
\sum_{lnr} \langle l^\prime n^\prime r^\prime|\hat{H}-E | lnr \rangle C^p_{lnr} &=
0. \label{eq:schrod_eq}
\end{align}
From this coefficient
matrix, we  obtain the new transformation matrix at site $p\pm 1 $
from the SVD in Eq. (\ref{eq:svd_relation}) (or equivalently, in the
density matrix formulation, from the eigenvectors of the density
matrix in Eq. (\ref{eq:dm_eigenvectors})).  If we  move
through the sites from left to right ($p \to p+1$) in a sweep, we successively
determine new $L^p$ matrices, while moving from right to left ($p \to p-1$) determines
new $R^p$ matrices. After the sites are traversed in one direction,
we traverse in the opposite direction  thus allowing improvement of all the $L^p$ and
$R^p$ matrices. (Of course, to initialise the procedure, requires some
starting guess for the  $L^p$ and $R^p$ matrices). This
is the basic method that is employed to optimise the
DMRG energy.

We usually depict the canonical representation at site $p$ in a block-configuration
diagram as shown in Fig. \ref{fig:blocking} consisting of a left block
of orbitals $1 \ldots p-1$, the site $p$
and a right block of orbitals $p+1 \ldots k$. Then, moving from one site
to another corresponds to moving from block-configuration to block-configuration, sweeping from
left-to-right and then right-to-left as shown in Fig. \ref{fig:sweep}.

So far  we have always been working within what is known
as the \textit{one-site} DMRG algorithm, since, as can be seen from
the block diagram in Fig. \ref{fig:blocking}, there is  only one site  between the left and right
blocks. However, in earlier formulations of the DMRG algorithm it was common to use the so-called two-site algorithm,
corresponding to the second block configuration in
Fig. \ref{fig:blocking}. Here the wavefunction  at site $p$ is written
in the renormalised product space as
\begin{equation}
|\Psi\rangle = \sum_{l n n^\prime r}
C^p_{lnn^\prime r}
 |l_{p-1} n_p n^\prime_{p+1} r_{p+1}\rangle
\end{equation}
where we see that two complete orbital Fock spaces $\{ n_p\}, \{n_{p+1}\}$
appear in the wavefunction expansion.
Unlike in the one-site configuration, we can only approximately
relate the canonical representations of the two-site wavefunctions at different
sites, and thus there is no single consistent DMRG wavefunction across
a two-site DMRG sweep, but
rather a whole family of DMRG wavefunctions, one at each site. 
Originally, the two-site algorithm was introduced to eliminate some
numerical problems associated with local minima when  optimising the DMRG wavefunction in the sweep
algorithm \cite{White1993, Chan2002, Chan2004}, but with the introduction of newer methods
which avoid such minima \cite{White2005}, the one-site formulation should now be viewed as preferred. 

\section{Evaluation of matrix elements}

\label{sec:matrix}
For completeness, we now outline briefly   how  the DMRG
wavefunction  allows the efficient evaluation of the  matrix
elements necessary to solve the Schrodinger equation in the
renormalised product basis (\ref{eq:schrod_eq}). 
We first note that any operator in the complete Fock space $\{ n_1 \} \otimes \{
n_2 \} \otimes \{ n_3 \} \otimes \ldots \otimes \{n_k\}$ can be
expressed as  a sum of products of ``local'' operators that each act individually in the
Fock space of a single site. For example, the unit operator $\hat{I}$
in the full Fock space may be considered as a single tensor product of
local unit operators
\begin{equation}
\hat{I} = \hat{I}^1 \otimes \hat{I}^2 \otimes \hat{I}^3 \otimes \ldots
\otimes \hat{I}^k
\end{equation}
where e.g. $\langle n_1 | \hat{I}^1 | n_1\rangle = \delta_{n_1
  n_1^\prime}$. To see how the   quantum chemistry Hamiltonian 
\begin{equation}
H=\sum_{ij} t_{ij} a^\dag_i a_j + \sum_{ijkl} v_{ijkl} a^\dag_i
a^\dag_j a_k a_l
\end{equation}
can be written as a sum of
products of local operators,
it is sufficient to show that the  creation and
annihilation operators can be expressed in this form.
Note that a single creation or annihilation operator  does not simply act in the Fock space of a single
orbital, because of the anticommutation relations between
operators. Instead, we write for   $a^\dag_i, a_i$
\begin{align}
a^\dag_i& =  \prod_{j<i} (-)^{n_{j}}  \otimes  P_i
a^\dag_i P_i, \\
a_i& =  \prod_{j<i} (-)^{n_{j}}  \otimes  P_i
a_i P_i.
\end{align}
Here the operator $\prod_{j<i} (-)^{n_{j}}$ formally keeps tracks of
the anticommutation, since if we consider e.g. $a_i$ acting
on a determinant, it counts the number of sign changes
involved in moving orbital $i$ to the front of the orbital string.  $P_i$ denotes the projection of operator onto
the $\{ n_i \}$ space alone. 

Given that all operators can be written as a sum of products of local
operators, we now examine how the matrix elements of a single product of
local operators are obtained.
Consider the product
\begin{equation}
\hat{O} = \hat{O}^1 \otimes \hat{O}^2 \otimes \ldots \hat{O}^k.
\end{equation}
In terms of the product basis $\{ |l_{p_1}n_pr_p\rangle \}$ of site
$p$, we can write (dropping the subscripts on the basis functions for simplicity)
\begin{align}
 \langle \lpp
\npp \rpp | \hat{O} | lnr\rangle& = \langle \lpp| \hat{O}^1 \otimes \ldots
\otimes \hat{O}^{p-1} | l\rangle \langle \npp | \hat{O}^p | n\rangle \langle
\rpp| \hat{O}^p \otimes \ldots \otimes \hat{O}^k |r\rangle \\
&= \langle \lpp| \hat{O}_L | l\rangle \langle \npp| \hat{O}^p | n\rangle
\langle \rpp| \hat{O}_R|r \rangle .
\end{align}
It is sufficient  to demonstrate how the matrix elements $\langle
l|\hat{O}_L|l^\prime\rangle$  are calculated
as those for $O_R$ are obtained in a similar manner. From the recursive definitions of the renormalised basis
functions $|l\rangle, |\lpp\rangle$ in
Eqs. (\ref{eq:left_transform_ip}), (\ref{eq:right_transform_ip}) we have
\begin{align}
\langle \lpp | \hat{O}_L |l\rangle& = \mathop{\sum_{n_1 \ldots n_p}}_{
 n_1^\prime \ldots n_p^\prime} [L^{n_1}][L^{n_2}] \ldots [L^{n_{p-1}}]
 \bigl(
O^1_{n_1 n_1^\prime} O^2_{n_2 n^\prime_2} \ldots O^{p-1}_{n_{p-1}
 n^\prime_{p-1}} 
\bigr)
[L^{n_1^\prime}][L^{n^\prime_2}] \ldots [L^{n^\prime_{p-1}}].
\end{align}
These multiple transformations may be efficiently  organised
into groups of two step procedures (corresponding to the familiar blocking and decimation steps of the
RG). Writing 
\begin{align}
O^1_{l_1 l^\prime_1} = \langle l_1|\hat{O}_1|
l_1^\prime \rangle = L^{n_1}_{l_1} O^1_{n_1 n^\prime_1} L^{n_1}_{l_1}
\end{align}
the blocking step corresponds to
\begin{align}
O^1_{l_1 l^\prime_1} \otimes O^2_{n_2 n^\prime_2}& \to \bigl(O^1
  O^2\bigr)_{l_1 n_1 l^\prime_1 n^\prime_1}
\end{align}
while the decimation corresponds to the  transformation into the renormalised basis ($\{ l_{1}
n_2\} \to \{ l_2 \}$)
\begin{align}
\sum_{l_1n_2l^\prime_1 n^\prime_2} L^{n_2}_{l_1 l_2} \bigl(O^1 O^2\bigr)_{l_1 n_2 l_1^\prime n_2^\prime}
L^{n_2^\prime}_{l_1^\prime  l_2^\prime}  \to \bigl(O^1 O^2\bigr)_{l_2 l_2^\prime}.
\end{align}
Each such transformation has the cost of a matrix multiplication i.e
$O(M^3)$, and  because of the  recursive structure of the
transformations, the complete matrix element $\langle \lpp
|O_L|l\rangle$  may be efficiently evaluated  as
a sequence of matrix products with a total cost  $O(M^3 k)$.

For complicated operators such as the 
quantum chemical Hamiltonian which consist of sums over many products of 
operators, it is clear that there are  intermediates which can be reused and
saved. For example, the matrix elements of $a^\dag_1
a^\dag_2 a_9 a_{10}$ and $a^\dag_1
a^\dag_2 a_4 a_5$  both involve  as an intermediate the renormalised
representation of $a^\dag_1 a^\dag_2$, which may  be stored and
reused. In practice, therefore, the optimal
implementation of the  DMRG algorithm in quantum chemistry requires an efficient
organisation of intermediates and this is primarily where most of the
complexity may be be found. The
interested reader is referred to the literature for further
details e.g. \cite{White1999, Mitrushenkov2001,
  Chan2002, Legeza2003dyn, Moritz2007, Chan2004}.

\section{Conclusions}

In this article we have attempted to introduce the Density Matrix
Renormalisation Group (DMRG)  primarily from the view that it
provides quantum chemistry with a new
kind of wavefunction ansatz. Consequently, we can analyse and
manipulate the ansatz in the way to which we are accustomed in
quantum chemistry. By examining its structure
we arrive at an intuitive understanding of the strengths of the DMRG
method e.g. in multireference problems, or in long molecules,
where it is a naturally local multireference approach. 
A striking  feature of the DMRG ansatz as compared to other
quantum chemical wavefunctions is the  
recursive  structure. This is the 
 connection between the DMRG wavefunction and the
 traditional language of
 the Renormalisation Group, and provides the  central mechanism
 behind the efficient evaluation of matrix elements in
 the method.

Traditionally quantum chemistry has  understood electronic structure
   in terms of the many-electron wavefunction. We hope that by  
  thinking about the DMRG  in this language,  it will not only become more accessible, but new
  possibilities will  arise  for  cross-fertilisation  between
   quantum chemical techniques and the Density Matrix
  Renormalisation Group. 

\label{sec:conclusions}

\section{Acknowledgments}
Garnet Kin-Lic Chan would like to acknowledge support from
Cornell University, the Cornell Center
for Materials Research (CCMR), the David and
Lucile Packard Foundation, the National Science
Foundation CAREER program CHE-0645380, the Alfred P. Sloan
Foundation, and the Department of Energy, Office of Science through
award DE-FG02-07ER46432.
Johannes Hachmann would like to acknowledge support provided by a Kekul\'{e} Fellowship of the Fond der
Chemischen Industrie. Eric Neuscamman would like to acknowledge
support provided by a National Science Foundation Graduate Research Fellowship.


\bibliographystyle{spphys}
\bibliography{dmrgwf_book}

\begin{thebibliography}{10}
\providecommand{\url}[1]{{#1}}
\providecommand{\urlprefix}{URL }
\expandafter\ifx\csname urlstyle\endcsname\relax
  \providecommand{\doi}[1]{DOI \discretionary{}{}{}#1}\else
  \providecommand{\doi}{DOI \discretionary{}{}{}\begingroup
  \urlstyle{rm}\Url}\fi

\bibitem{White1992}
S.R. White, Phys. Rev. Lett. \textbf{69}(19), 2863 (1992)

\bibitem{White1993}
S.R. White, Phys. Rev. B \textbf{48}(14), 10345 (1993)

\bibitem{Chan2003}
G.K.L. Chan, M.~Head-Gordon, J. Chem. Phys. \textbf{118}(19), 8551 (2003)

\bibitem{Chan2004b}
G.K.L. Chan, M.~K\'{a}llay, J.~Gauss, J. Chem. Phys. \textbf{121}(13), 6110
  (2004)

\bibitem{Hachmann2007}
J.~Hachmann, J.J. Dorando, M.~Avil\'{e}s, G.K.L. Chan, J. Chem. Phys.
  \textbf{127}(13), 134309  (2007)

\bibitem{Dorando2007}
J.J. Dorando, J.~Hachmann, G.K.L. Chan, J. Chem. Phys. \textbf{127}(8), 084109
  (2007)

\bibitem{Hachmann2006}
J.~Hachmann, W.~Cardoen, G.K.L. Chan, J. Chem. Phys. \textbf{125}(14), 144101
  (2006)

\bibitem{Wilson1975}
K.G. Wilson, Rev. Mod. Phys. \textbf{47}(4), 773 (1975)

\bibitem{Wilson1983}
K.G. Wilson, Rev. Mod. Phys. \textbf{55}(3), 583 (1983)

\bibitem{White1999}
S.R. White, R.L. Martin, J. Chem. Phys. \textbf{110}(9), 4127 (1999)

\bibitem{Mitrushenkov2001}
A.O. Mitrushenkov, G.~Fano, F.~Ortolani, R.~Linguerri, P.~Palmieri, J. Chem.
  Phys. \textbf{115}(15), 6815 (2001)

\bibitem{Chan2002}
G.K.L. Chan, M.~Head-Gordon, J. Chem. Phys. \textbf{116}(11), 4462 (2002)

\bibitem{Legeza2003dyn}
{\"O}.~Legeza, J.~R{\"o}der, B.A. Hess, Phys. Rev. B \textbf{67}(12), 125114
  (2003)

\bibitem{Moritz2007}
G.~Moritz, M.~Reiher, J. Chem. Phys. \textbf{126}(24), 244109 (2007)

\bibitem{Fannes1992}
M.~Fannes, B.~Nachtergaele, R.F. Werner, Comm. Math. Phys. \textbf{144}(3), 443
  (1992)

\bibitem{Fannes1994}
M.~Fannes, B.~Nachtergaele, R.F. Werner, J. Funct. Anal. \textbf{120}(2), 511
  (1994)

\bibitem{Ostlund1995}
S.~{\"O}stlund, S.~Rommer, Phys. Rev. Lett. \textbf{75}(19), 3537 (1995)

\bibitem{Rommer1997}
S.~Rommer, S.~{\"O}stlund, Phys. Rev. B \textbf{55}(4), 2164 (1997)

\bibitem{Verstraete2004mpdo}
F.~Verstraete, J.J. Garc\'{i}a-Ripoll, J.I. Cirac, Phys. Rev. Lett.
  \textbf{93}(20), 207204 (2004)

\bibitem{Verstraete2004pbc}
F.~Verstraete, D.~Porras, J.I. Cirac, Phys. Rev. Lett. \textbf{93}(22), 227205
  (2004)

\bibitem{Verstraete2004peps}
F.~Verstraete, J.I. Cirac, arXiv:cond-mat \textbf{0407066v1} (2004)

\bibitem{Perez-Garcia2007}
D.~P\'{e}rez-Garci\'{a}, F.~Verstraete, J.I. Cirac, M.M. Wolf, arXiv:quant-ph
  \textbf{0707.2260v1} (2007)

\bibitem{Schuch2007}
N.~Schuch, M.M. Wolf, F.~Verstraete, J.I. Cirac, Phys. Rev. Lett.
  \textbf{98}(14), 140506  (2007)

\bibitem{Murg2007}
V.~Murg, F.~Verstraete, J.I. Cirac, Phys. Rev. A \textbf{75}(3), 033605  (2007)

\bibitem{Verstraete2005}
F.~Verstraete, A.~Weichselbaum, U.~Schollw{\"o}ck, J.I. Cirac, J.~von Delft,
  arXiv:cond-mat \textbf{0504305v1} (2005)

\bibitem{White2004}
S.R. White, A.E. Feiguin, Phys. Rev. Lett. \textbf{93}(7), 076401 (2004)

\bibitem{Daley2004}
A.J. Daley, C.~Kollath, U.~Schollw{\"o}ck, G.~Vidal, J. Stat. Mech.: Theor.
  Exp. (04), P04005 (2004)

\bibitem{Vidal2004}
G.~Vidal, Phys. Rev. Lett. \textbf{93}(4), 040502  (2004)

\bibitem{Vidal2006}
G.~Vidal, arXiv:quant-ph \textbf{0610099v1} (2006)

\bibitem{Hallberg2003}
K.~Hallberg, in \emph{Theoretical Methods for Strongly Correlated Electrons},
  ed. by D.~S\'{e}n\'{e}chal, A.M. Tremblay, C.~Bourbonnais, CRM Series in
  Mathematical Physics (Springer, New York, 2003)

\bibitem{Hallberg2006}
K.A. Hallberg, Adv. Phys. \textbf{55}(5), 477 (2006)

\bibitem{Schollwock2005}
U.~Schollw{\"o}ck, Rev. Mod. Phys. \textbf{77}(1), 259 (2005)

\bibitem{Daul2000}
S.~Daul, I.~Ciofini, C.~Daul, S.R. White, Int. J. Quantum Chem. \textbf{79}(6),
  331 (2000)

\bibitem{Rissler2006}
J.~Rissler, R.M. Noack, S.R. White, Chem. Phys. \textbf{323}(2-3), 519 (2006)

\bibitem{Mitrushenkov2003}
A.O. Mitrushenkov, R.~Linguerri, P.~Palmieri, G.~Fano, J. Chem. Phys.
  \textbf{119}(8), 4148 (2003)

\bibitem{Mitrushenkov2003nort}
A.O. Mitrushenkov, G.~Fano, R.~Linguerri, P.~Palmieri, arXiv:cond-mat
  \textbf{0306058v1} (2003)

\bibitem{Chan2004}
G.K.L. Chan, J. Chem. Phys. \textbf{120}(7), 3172 (2004)

\bibitem{Chan2005}
G.K.L. Chan, T.~Van~Voorhis, J. Chem. Phys. \textbf{122}(20), 204101 (2005)

\bibitem{Legeza2003qie}
{\"O}.~Legeza, J.~S\'{o}lyom, Phys. Rev. B \textbf{68}(19), 195116 (2003)

\bibitem{Legeza2003lif}
{\"O}.~Legeza, J.~R{\"o}der, B.A. Hess, Mol. Phys. \textbf{101}(13), 2019
  (2003)

\bibitem{Legeza2004}
{\"O}.~Legeza, J.~S\'{o}lyom, Phys. Rev. B \textbf{70}(20), 205118  (2004)

\bibitem{Moritz2005orb}
G.~Moritz, B.A. Hess, M.~Reiher, J. Chem. Phys. \textbf{122}(2), 024107 (2005)

\bibitem{Moritz2005rel}
G.~Moritz, A.~Wolf, M.~Reiher, J. Chem. Phys. \textbf{123}(18), 184105 (2005)

\bibitem{Moritz2006}
G.~Moritz, M.~Reiher, J. Chem. Phys. \textbf{124}(3), 034103 (2006)

\bibitem{Zgid2008}
D.~Zgid, M.~Nooijen, J. Chem. Phys. In print

\bibitem{Ramasesha1997}
S.~Ramasesha, S.K. Pati, H.R. Krishnamurthy, Z.~Shuai, J.L. Br\'{e}das, Synth.
  Met. \textbf{85}(1-3), 1019 (1997)

\bibitem{Yaron1998}
D.~Yaron, E.E. Moore, Z.~Shuai, J.L. Br\'{e}das, J. Chem. Phys.
  \textbf{108}(17), 7451 (1998)

\bibitem{Shuai1998}
Z.~Shuai, J.L. Br\'{e}das, A.~Saxena, A.R. Bishop, J. Chem. Phys.
  \textbf{109}(6), 2549 (1998)

\bibitem{Fano1998}
G.~Fano, F.~Ortolani, L.~Ziosi, J. Chem. Phys. \textbf{108}(22), 9246 (1998)

\bibitem{Bendazzoli1999}
G.L. Bendazzoli, S.~Evangelisti, G.~Fano, F.~Ortolani, L.~Ziosi, J. Chem. Phys.
  \textbf{110}(2), 1277 (1999)

\bibitem{Raghu2002a}
C.~Raghu, Y.~Anusooya~Pati, S.~Ramasesha, Phys. Rev. B \textbf{65}(15), 155204
  (2002)

\bibitem{Raghu2002b}
C.~Raghu, Y.~Anusooya~Pati, S.~Ramasesha, Phys. Rev. B \textbf{66}(3), 035116
  (2002)

\bibitem{Dukelsky1998}
J.~Dukelsky, M.A. Mart\'{i}n-Delgado, T.~Nishino, G.~Sierra, Europhys. Lett.
  \textbf{43}(4), 457  (1998)

\bibitem{Verstraete2006}
F.~Verstraete, J.I. Cirac, Phys. Rev. B \textbf{73}(9), 094423  (2006)

\bibitem{Perez-Garcia2007mps}
D.~P\'{e}rez-Garci\'{a}, F.~Verstraete, M.M. Wolf, J.I. Cirac, Quant. Inf.
  Comp. \textbf{7}(5\&6), 401 (2007)

\bibitem{White2005}
S.R. White, Phys. Rev. B \textbf{72}(18), 180403 (2005)

\end{thebibliography}

\end{document}